# Near-Exact CASSCF-Level Geometry Optimization with a Large Active Space using Adaptive Sampling Configuration Interaction Self-Consistent Field Corrected with Second-Order Perturbation Theory (ASCI-SCF-PT2)

Jae Woo Park[*]

*Department of Chemistry, Chungbuk National University (CBNU), Cheongju 28644, Korea*

**Abstract**

An accurate description of electron correlation is one of the most challenging problems in quantum chemistry. The exact electron correlation can be obtained by means of full configuration interaction (FCI). A simple strategy for approximating FCI at a reduced computational cost is selected CI (SCI), which diagonalizes the Hamiltonian within only the chosen configuration space. Recovery of the contributions of the remaining configurations is possible with second-order perturbation theory. Here, we apply adaptive sampling configuration interaction (ASCI) combined with molecular orbital optimizations (ASCI-SCF) corrected with second-order perturbation theory (ASCI-SCF-PT2) for geometry optimization by implementing the analytical nuclear gradient algorithm for ASCI-PT2 with the *Z*-vector (Lagrangian) formalism. We demonstrate that for phenalenyl radicals and anthracene, optimized geometries and the number of unpaired electrons can be obtained at nearly the CASSCF accuracy by incorporating PT2 corrections and extrapolating them. We demonstrate the current algorithm's utility for optimizing the equilibrium geometries and electronic structures of 6-ring-fused polycyclic aromatic hydrocarbons and 4-periacene.

[*] E-mail: jaewoopark@cbnu.ac.kr

# 1. INTRODUCTION

In quantum chemistry, appropriate treatment of the correlations between electronic motions (electron correlation) is crucial to obtain accurate energies.[1-5] The perfect descriptions of electron correlation can be obtained via full configuration interaction (FCI), which diagonalizes the electronic Hamiltonian within the space of all possible electronic configurations (determinants). Unfortunately, the number of electronic configurations grows exponentially with the size of the orbitals. It is practically impossible to compute more than 22 electrons in 22 orbitals [i.e., (22$e$,22$o$)], even with a carefully tailored algorithm on modern massive parallel computational devices.[6]

Since the dawn of quantum chemistry, researchers have investigated many different approximate FCI methods. The density-matrix renormalization group (DMRG) method[7-9] has been shown to successfully approximate the FCI wave function in the form of matrix-product states (MPS).[10,11] The DMRG method is considered a *de facto* standard when FCI is not possible. The variational two-electron reduced density matrix (v2RDM) method, which employs the 2-particle reduced density as a variational parameter, can treat the electron correlation to an impressive degree of accuracy.[12] The FCI quantum Monte Carlo (FCIQMC) method, based on quantum Monte Carlo propagation, has been very successful in terms of accuracy and robustness, even for highly correlated systems such as solids.[13-16] There are also incremental approximations to FCI theory.[17-26]

Another simple approach is the selected CI (SCI) concept, which includes representative determinants in the wave function. The oldest SCI algorithm is probably the Configuration Interaction using a Perturbative Selection made Iteratively (CIPSI) method.[27] There are many ways to sample determinants. For example, the Monte Carlo sampling by Greer,[28-30] the determinant-driven CIPSI,[31-34] heat-bath CI (HCI),[35,36] adaptive CI (ACI),[37] iterative CI[38] with selection,[39] and adaptive sampling CI (ASCI)[40,41] have been developed in the last two decades. The ASCI method is a cost-effective and deterministic way to sample determinants without introducing any randomness during the procedure.[41-43] Recently, analogous methods were also employed to treat vibrational problems.[44-50] For more details, the readers can refer to Eriksen's recent comprehensive perspective on FCI solvers.[51]

One can embed a highly correlated system (a set of orbitals) into a mean-field system. We denote these orbitals as "active," and this embedding results in the complete active space self-consistent field (CASSCF) method,[52-55] which was also known as fully optimized reactive space (FORS).[54] The FCI calculation performed in the active space is complete active space CI (CASCI). The computational demand for CASCI can be prohibitive when the active space is large. A useful approximation is to impose a restriction on the orbitals' occupancies, which results in the restricted active space (RAS)[56,57] and generalized active space (GAS)[58-60] concepts. Of course, the approximate FCI methods can be used in place of CASCI. While the CIPSI method was combined with multiconfigurational SCF in 1986,[61] approximate CASSCF methods were developed for DMRG,[8] v2RDM,[62] FCIQMC,[63-65] and HCI[66] theories in the recent two decades. For ASCI, the ASCI-SCF method was developed and compared with CASSCF.[67,68]

The SCI methods include only some of the configurations in the variational Hamiltonian. The rest of the energy can be corrected for by applying perturbation theory,[27,40,42,66-69] as done by the CIPSI method.[27] This correction can be utilized in the SCI method embedded in the mean field as well. Usually, the Epstein–Nesbet partitioning of the electronic Hamiltonian is exploited for this purpose.[70,71] The HCISCF method includes a PT2 correction for the orbital optimizations to obtain orbitals at the near-CASSCF level.[66] In contrast, only variational SCI energies are applied in the orbital optimizations in the ASCI-SCF methods, and the PT2 contributions are evaluated later.[67,68] In both cases, obtaining energies comparable to those of CASSCF is possible.[22,23]

Geometry optimizations and dynamics simulations are among the most important applications of quantum chemistry calculations. In terms of computational cost and accuracy, having an analytical nuclear gradient is advantageous for such simulations. Recently, we developed analytical gradient methods for variational ASCI-SCF and applied this method to optimize the molecular geometries of periacenes and polyacenes.[68] The analytical gradient used in Ref. 68 did not include the PT2 correction to the analytical gradient. Therefore, improvement in the quality of the optimized geometry by incorporating the PT2 correction is warranted.

In this work, we formulate an analytical gradient method for perturbatively corrected ASCI-SCF, which we call ASCI-SCF-PT2, by implementing the response equation (or the so-called $Z$-vector equation),[72-76] as the ASCI-SCF-PT2 energy is not variational with respect to the orbital rotations and the CI coefficients. We find that for phenalenyl radicals and anthracene, the quality of the ASCI-SCF-PT2 optimized geometries is comparable to that of CASSCF-generated geometries, while the improvements in energies are minor. Moreover, by extrapolating the ASCI-SCF-PT2 nuclear gradient and number of unpaired electrons, near-exact CASSCF-level geometries and radical indices are obtained. We demonstrate the algorithm's utility for optimizing the ground-state geometries of 6-ring polycyclic aromatic hydrocarbons (PAHs) and 4-periacene.

## 2. THEORY

This section first reviews the ASCI algorithm and the second-order perturbation theory correction to the ASCI energy. Then, we introduce the analytical nuclear gradient theory for the ASCI-SCF-PT2 energy. The indices $I, J, K, \ldots, T, U, V, \ldots, i, j, k, \ldots, r, s, t, \ldots,$ and $x, y, z, \ldots,$ denote electronic configurations in the SCI space, electronic configurations out of the SCI space, doubly occupied (closed) orbitals, active orbitals, and general orbitals, respectively.

**Adaptive Sampling CI Algorithm.** In the ASCI method,[40,41,67] the wave function is

$$|\Psi\rangle = \sum_I^{S_D} c_I |I\rangle, \quad (1)$$

where $S_D$ includes representative electronic configurations sampled as follows. First, the Hartree–Fock (HF) ground-state determinant, $|0\rangle$, is employed as an initial guess. For all determinants that interact with the HF determinant, the perturbative amplitudes are computed as

$$A_T = \frac{\langle 0|\hat{H}|T\rangle}{E_{\text{HF}} - H_{TT}}. \quad (2)$$

Here, $E_{\text{HF}}$ is the HF energy, and $H_{TT}$ is the diagonal element of the Hamiltonian for determinant $T$. $N_{\text{tdet}}$ "target" determinants are selected based on the perturbative amplitudes. Then, the Hamiltonian is constructed and diagonalized in the target space. The $N_{\text{cdet}}$ "core" determinants are selected based on the eigenvectors (CI coefficients). Then, the singly and doubly excited determinants are again generated from the core determinants, with perturbative amplitudes of

$$A_T = \frac{\sum_I^{\text{core}} H_{TI} c_I}{E - H_{TT}}, \quad (3)$$

and then, the target determinants are again selected. This procedure is iteratively repeated until convergence is reached in terms of the ground-state energy. In our implementation, the default threshold is 1 $\mu E_\text{h}$. This target determinant selection strategy can also be applied for computing excited states, while one would utilize a state-by-state bootstrap method.[40] We embed the ASCI orbital space as the active space to establish the ASCI-SCF method. The second-order augmented

Hessian (AH) orbital optimization scheme was used in our implementation, while we updated the list of the target determinants every 10 macroiterations ($N_{\text{macro}} = 10$).

**Second-Order Perturbation Theory**. The Epstein–Nesbet perturbation theory (ENPT)[70,71] corrects the ASCI energy to account for the contributions from the determinants outside the SCI space.[27,36,37,42,66,69,77] The zeroth-order Hamiltonian in the ENPT is

$$\hat{H}^{(0)} = \hat{P}\hat{H}\hat{P} + \sum_{T} \hat{Q}_T H_{TT} \hat{Q}_T, \tag{4}$$

where $\hat{P}$ and $\hat{Q}_T$ are projectors onto the SCI subspace and a determinant $T$ in the rest of the FCI space, respectively. In variational perturbation theory, PT2 is reformulated as a minimization of the Hylleraas functional[78,79]

$$E_{\text{PT2}} = 2\langle \Psi^{(1)} | \hat{H} | \Psi^{(0)} \rangle + \langle \Psi^{(1)} | \hat{H}^{(0)} - E^{(0)} | \Psi^{(1)} \rangle, \tag{5}$$

where $E^{(0)}$ is the ASCI-SCF energy. Here, we parameterize the first-order correction to the wavefunction as

$$|\Psi^{(1)}\rangle = \sum_T A_T |T\rangle. \tag{6}$$

Solving the equation

$$\frac{\partial E_{\text{PT2}}}{\partial A_T} = 0 \tag{7}$$

yields the amplitude $A_T$ as

$$A_T = \frac{\langle \Psi^{(0)} | \hat{H} | T \rangle}{E - H_{TT}} = \sum_I \frac{c_I \langle I | \hat{H} | T \rangle}{E - H_{TT}}. \tag{8}$$

The final second-order energy with the ENPT2 is

$$E_{\text{PT2}} = \sum_T \frac{|\langle \Psi^{(0)} | \hat{H} | T \rangle|^2}{E - H_{TT}}$$
$$= \sum_T A_T \langle T | \hat{H} | \Psi^{(0)} \rangle = \sum_T \sum_I A_T c_I H_{TI} \qquad (9)$$

One should evaluate the perturbative correction by looping over the determinants $T$. We have used the algorithm based on triplet constraints, which groups $T$ based on their three highest-occupied alpha orbitals.[42] Compared to the naïve algorithm without such grouping, this process dramatically reduces the memory requirement and cost for sorting $T$.[39,42] In our implementation, the ASCI-SCF energy is corrected with the ENPT to arrive at the ASCI-SCF-PT2 energy.

**Analytical Gradient Theory: ASCI Lagrangian.** Instead of directly differentiating ASCI-SCF-PT2 energy, we can derive the analytical gradient more elegantly with the Lagrangian formalism.[72-76,80-85] The Lagrangian is

$$\mathcal{L} = E_{\text{ASCI-PT2}} + \sum_{xy} Z_{xy} \left( \frac{\partial E^{(0)}}{\partial U_{xy}} - \frac{\partial E^{(0)}}{\partial U_{yx}} \right) - \frac{1}{2} \sum_{xy} X_{xy} \left( S_{xy} - \delta_{xy} \right)$$
$$+ \left[ \sum_I z_I \left( \frac{\partial E^{(0)}}{\partial c_I} - E^{(0)} c_I \right) - \frac{1}{2} x \left( \sum_{IJ} c_I c_J \langle I | J \rangle - 1 \right) \right], \qquad (10)$$

which is stationary with respect to the variations in **C** and **c**, with suitable values of Lagrange multipliers **Z**, **z**, and **X**. We treat the convergence criteria for ASCI-PT2 as constraints. The matrices **Z** and **X** are antisymmetric and symmetric, respectively, from the form of these constraints. Also, **z** is set orthogonal to **c**. Instead of directly using **C** as the variational parameter, we use a unitary transformation matrix **U**, where

$$\mathbf{C} = \mathbf{C}^{(0)} \mathbf{U} = \mathbf{C}^{(0)} \exp(\boldsymbol{\kappa}), \qquad (11)$$

and $\boldsymbol{\kappa}$ is an antisymmetric matrix. This Lagrangian is stationary when

$$\frac{\partial \mathcal{L}}{\partial U_{xy}} = 0, \tag{12}$$

$$\frac{\partial \mathcal{L}}{\partial c_I} = 0. \tag{13}$$

When we define

$$Y_{xy} = \frac{\partial E_{\text{ASCI-PT2}}}{\partial U_{xy}}, \tag{14}$$

$$y_I = \frac{\partial E_{\text{ASCI-PT2}}}{\partial c_I}, \tag{15}$$

while $E_{\text{ASCI-PT2}}$ is the total energy that includes the ASCI energy and perturbative corrections, $E_{\text{ASCI-PT2}} = E_{\text{ASCI-SCF}} + E_{\text{PT2}}$. The Z-vector equation is then basically the same as that for CASSCF but with only limited dimensions in CI space. The explicit expression for the Z-vector equation, which is a coupled form of Eqs. (12) and (13), is

$$\begin{pmatrix} \dfrac{d^2 E}{d\boldsymbol{\kappa} d\boldsymbol{\kappa}} & \dfrac{d^2 E}{d\boldsymbol{\kappa} d\mathbf{c}} \\ \dfrac{d^2 E}{d\mathbf{c} d\boldsymbol{\kappa}} & \dfrac{d^2 E}{d\mathbf{c} d\mathbf{c}} \end{pmatrix} \begin{pmatrix} \mathbf{Z} \\ \mathbf{z} \end{pmatrix} = -\begin{pmatrix} \mathbf{Y} - \mathbf{Y}^\dagger \\ \mathbf{y} \end{pmatrix}. \tag{16}$$

Here, we have exploited the symmetricity of **X** to remove it from the equation. One can solve this equation using coupled[72-76] or uncoupled schemes.[80] We have confirmed that both techniques resulted in the same gradients.

**Source Terms for ASCI-PT2.** The source terms **Y** and **y** should be evaluated to solve the Z-vector equation. The Hylleraas functional is stationary with respect to variations in $A_T$. The Hylleraas functional in terms of **A**, **c**, and **H** is

$$E_{\text{ASCI-PT2}} = E^{(0)} + E_{\text{PT2}}$$
$$= E^{(0)} + 2\sum_{TI} A_T c_I H_{TI} + \sum_T A_T^2 (H_{TT} - \sum_{IJ} c_I c_J H_{IJ}). \tag{17}$$

Here, the zeroth-order energy (ASCI-SCF energy) is

$$E^{(0)} = \sum_{IJ} H_{IJ} c_I c_J = \sum_{xy} h_{xy} d_{xy}^{(0)} + \sum_{xyzw} (xy \mid zw) D_{xy,zw}^{(0)} \tag{18}$$

The mathematical derivations can be made simpler by defining the density matrices. The total density matrices are

$$\mathbf{d} = \mathbf{d}^{(0)} + \mathbf{d}^{(1)} + \mathbf{d}^{(2)}, \tag{19}$$

$$\mathbf{D} = \mathbf{D}^{(0)} + \mathbf{D}^{(1)} + \mathbf{D}^{(2)}, \tag{20}$$

where the superscript ($n$) indicates that the matrices include the $n$-th-order term of the amplitudes. The total ASCI-PT2 energy is

$$E_{\text{ASCI-PT2}} = \sum_{rs} h_{rs} d_{rs} + \sum_{rstu} (rs \mid tu) D_{rs,tu}. \tag{21}$$

The density matrices are then

$$d_{rs}^{(1)} = \sum_{TI} A_T c_I \left[ \frac{\partial H_{TI}}{\partial h_{rs}} + \frac{\partial H_{IT}}{\partial h_{rs}} \right], \tag{22}$$

$$D_{rs,tu}^{(1)} = \sum_{TI} A_T c_I \left[ \frac{\partial H_{TI}}{\partial (rs \mid tu)} + \frac{\partial H_{IT}}{\partial (rs \mid tu)} \right], \tag{23}$$

$$d_{rs}^{(2)} = \sum_T A_T^2 \frac{\partial H_{TT}}{\partial h_{rs}} - d_{rs}^{(0)} N, \tag{24}$$

$$D_{rs,tu}^{(2)} = \sum_T A_T^2 \frac{\partial H_{TT}}{\partial (rs \mid tu)} - D_{rs,tu}^{(0)} N, \tag{25}$$

where we have defined the perturbative norm, $N = \sum_T A_T^2$. We symmetrize the first-order density matrices. Should there be closed (doubly occupied) orbitals, the core Fock integrals replace the one-electron integrals. With these density matrices, the orbital gradient $\mathbf{Y}$ is obtained in the same way that the orbital gradients in the CASSCF calculation are evaluated. The only difference is that the zeroth-order density matrix is replaced with the total density matrix. The CI derivative is simply

$$y_I = 2\left[\sum_T A_T H_{TI} + (1-N)\sum_J H_{IJ} c_J\right] \tag{26}$$
$$= 2\left[\sum_T A_T H_{TI} + (1-N) c_I E^{(0)}\right],$$

and as **z** is taken to be normal with **c**, only the first term is needed in the working implementation.

**Final Gradient Evaluation and Extrapolation.** After the Z-vector equation is solved, the final gradient is evaluated as

$$\frac{dE_{\text{ASCI-PT2}}}{d\mathbf{R}} = \frac{\partial \mathcal{L}}{\partial \mathbf{R}} = \sum_{\mu\nu} \frac{dh_{\mu\nu}}{d\mathbf{R}} d_{\mu\nu}^{\text{eff}} + \sum_{\mu\nu\lambda\sigma} \frac{d(\mu\nu|\lambda\sigma)}{d\mathbf{R}} D_{\mu\nu\lambda\sigma}^{\text{eff}} + \sum_{\mu\nu} \frac{dS_{\mu\nu}}{d\mathbf{R}} X_{\mu\nu}, \tag{27}$$

where the partial differentiation with respect to the nuclear coordinate **R** means that the Lagrangian is evaluated with the Cartesian integral derivatives. One then evaluates the effective (relaxed) densities $\mathbf{d}^{\text{eff}}$ and $\mathbf{D}^{\text{eff}}$ in the AO basis as

$$\mathbf{d}^{\text{eff}} = \mathbf{C}\left[\mathbf{d} + \bar{\mathbf{d}} + \mathbf{Z}\mathbf{d}^{(0)} + \mathbf{d}^{(0)}\mathbf{Z}^?\right]\mathbf{C}^\dagger, \tag{28}$$

$$D_{\mu\nu\lambda\sigma}^{\text{eff}} = \sum_{zw} D_{\nu\sigma}^{\text{eff},zw} C_{\mu z} C_{\lambda w}, \tag{29}$$

$$\mathbf{D}^{\text{eff},zw} = \mathbf{C}\left[\mathbf{D}^{zw} + \frac{1}{2}\bar{\mathbf{Q}}^{zw} + \mathbf{Z}\mathbf{Q}^{zw} + \mathbf{Q}^{zw}\mathbf{Z}^?\right]\mathbf{C}^\dagger, \tag{30}$$

where

$$\bar{\mathbf{d}} = \frac{1}{2}\sum_I z_I \left\langle I \left| \hat{E}_{xy} + \hat{E}_{yx} \right| \Psi^{(0)} \right\rangle, \tag{31}$$

$$\bar{\mathbf{Q}}_{xy}^{zw} = \frac{1}{2}\sum_I z_I \left\langle I \left| \hat{E}_{xz,yw} + \hat{E}_{zx,yw} \right| \Psi^{(0)} \right\rangle, \tag{32}$$

$D_{xy}^{zw} = D_{xz,yw}$, and $Q_{xy}^{zw} = D_{xz,yw}^{(0)}$. These expressions are the same as those in Refs. 80 and 86.

One can extrapolate the PT2-corrected selected CI energy with respect to the PT2 correction to obtain the near-exact CASSCF energy.[39,40,42,43,67,87] Simple linear regressions are usually sufficient for this purpose, and the $R^2$ values for the fitting exceed 0.99 in almost all cases we have tested. We found that it is also possible to extrapolate the ASCI-SCF-PT2 nuclear gradients at each nuclear coordinate with respect to the PT2 correction to the point where the PT2 correction is zero to achieve near-exact CASSCF-level gradients. To this end, we compute the ASCI-SCF-PT2 analytical gradients for each $N_{\text{tdet}}$ (usually four to five points) and perform linear regressions for $3N_{\text{atom}}$ components of nuclear gradients to obtain the final gradient. At each extrapolation point, we perform the ASCI-SCF orbital optimization, but the iterative update of the list of the determinant is performed only at the point with the largest $N_{\text{tdet}}$ (i.e., $N_{\text{macro}} = \infty$ for each supermacroiteration in the ASCI-SCF optimization with smaller $N_{\text{tdet}}$), as this approximation does not affect the efficiency to a significant degree. The linear fitting cost is negligible, and the significant computational burden for this procedure is the performance of the ASCI-SCF-PT2 correction at each $N_{\text{tdet}}$. We will denote the extrapolated ASCI-SCF-PT2 gradient as the "ASCI-SCF-PT2+X" gradient for the sake of brevity.

## 3. NUMERICAL EXAMPLES

We have linked the programs for evaluating ASCI-SCF-PT2 density matrices and CI derivatives and solving the ASCI-SCF Z-vector equation to the program package BAGEL.[88] We computed all two-electron integrals with density fitting (DF) approximations. We used the cc-pVDZ basis set[89] and its corresponding JKFIT basis for DF approximation[90] unless otherwise mentioned.

**Comparison with CASSCF: Optimized geometries.** To assess the improvement offered by the PT2 corrections, we used phenalenyl radicals and anthracene (Figure 1) as test molecules. In the ASCI-SCF-PT2+X calculations, we employed five data points with $N_{tdet}$ = 20000, 40000, 60000, 80000, and 100000. The errors in the energies and optimized geometries, compared to the CASSCF (FCI) references, are shown in Table 1.

First, we can notice that the improvements in the energy are rather minor compared to the ASCI-SCF-PT2 energy at the ASCI-SCF geometries, particularly when $N_{tdet}$ is large. The largest improvement is 0.168 m$E_h$ (which is ~0.4 kJ/mol) for triplet anthracene with $N_{tdet}$ = 1000, while the improvement is below 0.001 m$E_h$ for (13$e$,13$o$) phenalenyl radicals with $N_{tdet}$ = 100000. In other words, to obtain an accurate ASCI-SCF-PT2 energy, it is sufficient to use the ASCI-SCF gradient in geometry optimization and perform a perturbative correction at the optimized geometry, particularly when $N_{tdet}$ is large. We note that this argument still holds with larger active spaces tested in this work–the differences between the ASCI-SCF-PT2 energies at the ASCI-SCF-PT2 and ASCI-SCF geometries were below 0.5 m$E_h$ in all cases. When ASCI-SCF-PT2+X gradients were used in optimizations, the resulting errors were within the 0.2 m$E_h$ range of the CASSCF value, which is a very reliable estimate considering possible extrapolation errors.

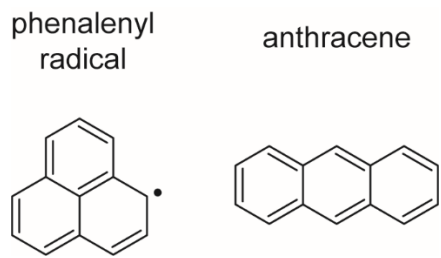

**Figure 1.** Structures of the molecules used for tests against CASSCF results.

**Table 1.** The errors in the energy (in m$E_h$) and geometries (Å, root-mean-squared distance) between the ASCI-SCF and ASCI-SCF-PT2 optimized energy and geometries with respect to the CASSCF reference values.

| | active space | $N_{cdet}$ | $N_{tdet}$ | $N_{CASCI}$[a] | $N_{tdet}/N_{CASCI}$ (%) | Errors in optimized energy (m$E_h$) | | | | RMS distance from CASSCF optimized geometry (Å) | | |
|---|---|---|---|---|---|---|---|---|---|---|---|---|
| | | | | | | ASCI-SCF | ASCI-SCF-PT2 at ASCI-SCF | ASCI-SCF-PT2 opt | ASCI-SCF-PT2+X opt | ASCI-SCF | ASCI-SCF-PT2 | ASCI-SCF-PT2+X |
| Phenalenyl radical (doublet) | (13$e$,13$o$) | 200 | 1000 | 2944656 | 0.03 | 23.111 | 7.423 | 7.381 | | 0.005 | 0.003 | |
| | | 200 | 10000 | 2944656 | 0.34 | 9.703 | 2.764 | 2.751 | -0.067 | 0.003 | 0.001 | 0.000 |
| | | 200 | 100000 | 2944656 | 3.40 | 1.414 | 0.341 | 0.341 | | 0.000 | 0.000 | |
| Anthracene (singlet) | (14$e$,14$o$) | 200 | 1000 | 11778624 | 0.01 | 24.892 | 8.368 | 8.333 | | 0.007 | 0.003 | |
| | | 200 | 10000 | 11778624 | 0.08 | 6.741 | 2.292 | 2.271 | -0.134 | 0.002 | 0.001 | 0.000 |
| | | 1000 | 100000 | 11778624 | 0.85 | 2.023 | 0.543 | 0.527 | | 0.001 | 0.000 | |
| Anthracene (triplet) | (14$e$,14$o$) | 200 | 1000 | 11778624 | 0.01 | 26.459 | 10.281 | 10.113 | | 0.006 | 0.003 | |
| | | 200 | 10000 | 11778624 | 0.08 | 10.184 | 3.677 | 3.620 | -0.140 | 0.003 | 0.002 | 0.000 |
| | | 1000 | 100000 | 11778624 | 0.85 | 1.982 | 0.588 | 0.586 | | 0.001 | 0.000 | |

[a] Number of determinants in CASCI space.

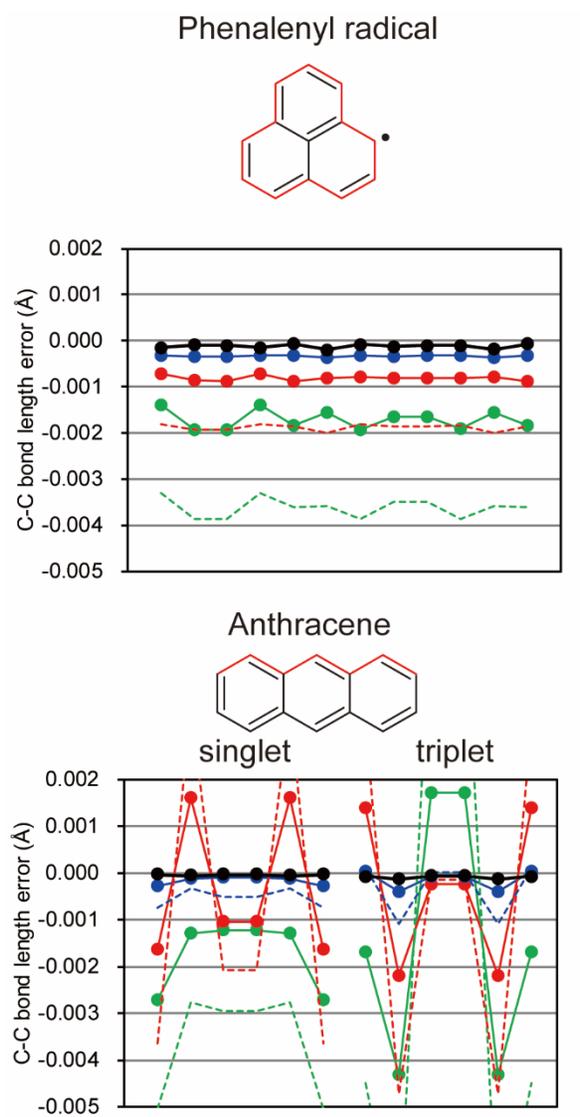

**Figure 2.** Errors in bond lengths with respect to the CASSCF values in the phenalenyl radical and anthracene for the geometries optimized with ASCI-SCF (dashed), ASCI-SCF-PT2 (thin solid) with $N_{tdet} = 1000$ (green), 10000 (red), 100000 (blue), and ASCI-SCF-PT2+X (thick black).

The ASCI-SCF-PT2 geometries are two times closer to the CASSCF geometries than the ASCI-SCF geometries in terms of root-mean-squared distances. The geometries obtained with extrapolated gradients are within a distance of 0.0003 Å from the CASSCF geometries. The deviations in the bond lengths of edge C–C bonds in phenalenyl radicals and anthracene from the CASSCF values are shown in Figure 2. In both systems, the errors are significantly reduced by increasing $N_{tdet}$. Without PT2 corrections, the maximum error decreases from 0.008 Å to 0.001 Å (anthracene) and from 0.004 Å to 0.000 4 Å (phenalenyl radical) when $N_{tdet}$ increases from 1000 to 100000. By considering the PT2 corrections, the absolute errors decrease by almost half. Including ~1% of complete electronic configurations in the ASCI-SCF-PT2 optimizations results in bond length deviations below 0.000 5 Å. The geometries with the ASCI-SCF-PT2+X gradients exhibit errors below 0.000 2 Å. Overall, these test results show that the ASCI-SCF-PT2 geometry optimizations can yield reasonable estimates of CASSCF-optimized geometries without the computational burden required for FCI diagonalization. Additionally, with linear extrapolation (ASCI-SCF-PT2+X), near-exact CASSCF-level geometries can be obtained.

**Comparison with CASSCF: Number of Unpaired Electrons.** The PT2 energy correction naturally improves the wavefunctions to the first order. In terms of RDMs, these corrections are given by Eqs. 19 and 20. Moreover, in analytical gradient theory, we make the Lagrangian (Eq. 10) stationary with respect to the orbital rotations by solving the Z-vector equation (Eqs. 12 and 13). In the sense of invariance, the Lagrangian is physically equivalent to the ASCI-SCF energy. The effective (relaxed) densities (Eq. 27), the RDMs for the Lagrangian, can be employed to analyze the electronic structures with a perturbative correction. This is a well-known strategy for examining electronic distributions in correlated methods.[91]

**Table 2.** The numbers of unpaired electrons, $N_U$, evaluated with effective (relaxed), unrelaxed ASCI-SCF-PT2, and ASCI-SCF densities and their errors with respect to the CASSCF results.

| $N_{tdet}$ | Density | phenalenyl radical (doublet) | | anthracene (singlet) | | anthracene (triplet) | |
|---|---|---|---|---|---|---|---|
| | | $N_U$ | error | $N_U$ | error | $N_U$ | error |
| 100000 | Effective | 1.351 | 0.008 | 0.427 | 0.013 | 2.357 | 0.014 |
| | ASCI-SCF-PT2 | 1.338 | 0.021 | 0.408 | 0.033 | 2.339 | 0.032 |
| | ASCI-SCF | 1.335 | 0.024 | 0.404 | 0.036 | 2.335 | 0.035 |
| 10000 | Effective | 1.315 | 0.044 | 0.379 | 0.062 | 2.304 | 0.067 |
| | ASCI-SCF-PT2 | 1.274 | 0.085 | 0.330 | 0.110 | 2.260 | 0.111 |
| | ASCI-SCF | 1.259 | 0.100 | 0.320 | 0.121 | 2.247 | 0.124 |
| 1000 | Effective | 1.259 | 0.099 | 0.312 | 0.128 | 2.237 | 0.134 |
| | ASCI-SCF-PT2 | 1.197 | 0.162 | 0.235 | 0.206 | 2.173 | 0.198 |
| | ASCI-SCF | 1.169 | 0.190 | 0.205 | 0.235 | 2.147 | 0.224 |
| CASSCF | | 1.359 | | 0.441 | | 2.371 | |
| Extrapolated | | 1.357 | | 0.442 | | 2.370 | |

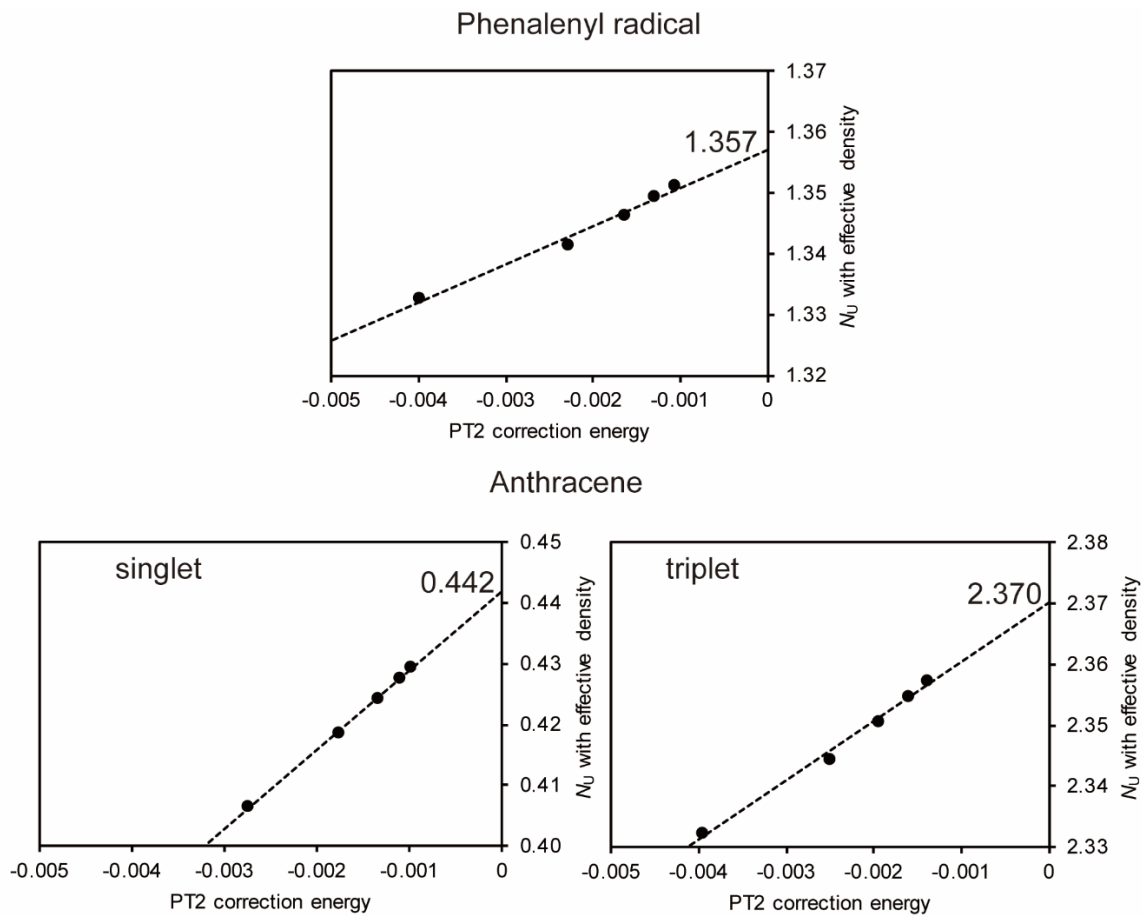

**Figure 3.** Results of linear extrapolation of the number of unpaired electrons ($N_U$) with respect to the PT2 correction energy in phenalenyl radicals (doublet) and anthracene (singlet and doublet). All $N_U$ values are evaluated using relaxed densities.

As an example, we evaluated the number of unpaired electrons with ASCI-SCF density and ASCI-SCF-PT2 densities (both unrelaxed and relaxed) in phenalenyl radicals and anthracene. We used the nonlinear model of Head-Gordon.[92] A quartic function of the natural occupancy $n_i$ of each natural orbital

$$N_U(i) = n_i^2(2-n_i)^2 \qquad (33)$$

is used to count the unpaired electrons in the system ($N_U$) by summing over the natural orbitals, $N_U = \sum_i N_U(i)$. By back-transforming a matrix with diagonal elements of $N_U(i)$ into the atomic orbital basis, these unpaired electrons can be assigned to each atom as well.

The numbers of unpaired electrons, compared to the CASSCF values, are shown in Table 2. As expected, the errors were reduced by increasing $N_{tdet}$: With the ASCI-SCF density, the errors were reduced by 0.16 to 0.20 when we increased $N_{tdet}$ from 1000 to 100000. The unrelaxed ASCI-SCF-PT2 densities increased $N_U$, but to a minor degree. Inclusion of the Z-vector contributions in densities resulted in $N_U$ values closer to the CASSCF values. In the largest calculations, $N_U$ differed from the reference value by only ~0.01. The $N_U$ values evaluated with ASCI-SCF and ASCI-SCF-PT2 were always smaller than the CASSCF values in all tested cases, which implies that the FCI space outside the ASCI space contributes to an increase in the unpaired electrons (which is very physical and natural).

We performed linear extrapolation of $N_U$ computed with the ASCI-SCF-PT2 relaxed density at the geometries optimized with $N_{tdet} = 100000$. We have included five data points ($N_{tdet}$ = 20000, 40000, 60000, and 80000) in these extrapolations. The results of such extrapolations are shown in Figure 3. The $N_U$ values depend almost linearly on the PT2 correction energies, and the differences between the resulting estimates and CASSCF values were below 0.002 for all tested cases. We note that the $N_U$ value from the ASCI-SCF-PT2 unrelaxed densities does not have this property. This finding shows that despite the additional nonlinearity introduced by the orbital optimizations in ASCI-SCF, extrapolation also works for $N_U$. This also suggests that this strategy could be used to compute reliable estimates of various electronic structure measures at the CASSCF level.

The extrapolations of both analytical gradients and $N_U$ were quite successful. We hypothesize that the density matrices in the atomic orbital basis can be the target function for extrapolation. Unfortunately, it is not easy to satisfy the physical properties of the density matrices with linear extrapolation, and thus, we leave it as a target for future investigations.

**Demonstration for 6-ring-fused PAHs.** Next, we demonstrate the ASCI-SCF-PT2 analytical gradient method for the group of 6-ring-fused PAHs shown in Figure 4. Among these molecules, uthrene and triangulene are predicted to have triplet ground states, according to Ovchinnikov's rule,[93] DFT,[94,95] MR-CISD, and MR-AQCC calculations.[95] Here, we perform ASCI-SCF-PT2 geometry optimizations for these molecules by employing full $\pi$-valence active space, which results in an active space ranging from (22$e$,22$o$) for triangulene to (26$e$, 26$o$) for hexacene and fulminene. We used $N_{tdet}$ = 500000 and $N_{cdet}$ = 1000 for the ASCI-SCF and ASCI-SCF-PT2 optimizations and $N_{tdet}$ = 100000, 200000, 300000, 400000, and 500000 for the extrapolation to evaluate the ASCI-SCF-PT2+X gradient.

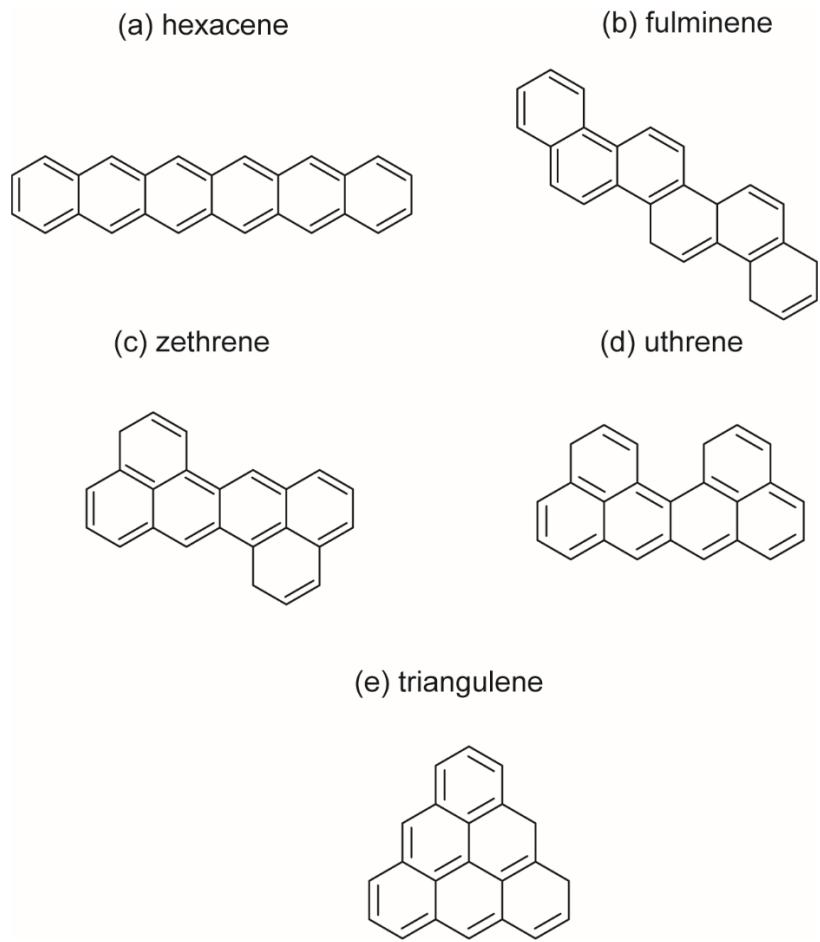

**Figure 4.** Molecular structures of 6-ring-fused PAHs.

**Table 3.** The adiabatic singlet–triplet gaps and the numbers of unpaired electrons ($N_U$) in 6-ring PAHs, computed with ASCI-SCF, ASCI-SCF-PT2, and extrapolated ASCI-SCF-PT2.

| | Singlet – triplet gaps ($\Delta E_{ST} = E_T - E_S$) (kcal/mol) | | | $N_U$ | | | |
|---|---|---|---|---|---|---|---|
| | $N_{tdet}$ = 500000 | | ASCI-SCF-PT2+X | $N_{tdet}$ = 500000 | | Extrapolated | |
| | ASCI-SCF | ASCI-SCF-PT2 | | singlet | triplet | singlet | triplet |
| hexacene | 19.30 | 18.96 | 19.34 | 0.73 | 2.49 | 1.01 | 2.65 |
| fulminene | 58.43 | 60.01 | 59.39 | 0.43 | 2.50 | 0.59 | 2.67 |
| zethrene | 28.82 | 25.94 | 22.19 | 0.57 | 2.44 | 0.75 | 2.59 |
| uthrene | −10.69 | −10.87 | −9.06 | 2.38 | 2.44 | 2.52 | 2.56 |
| triangulene | −12.36 | −13.48 | −13.48 | 2.41 | 2.46 | 2.48 | 2.56 |

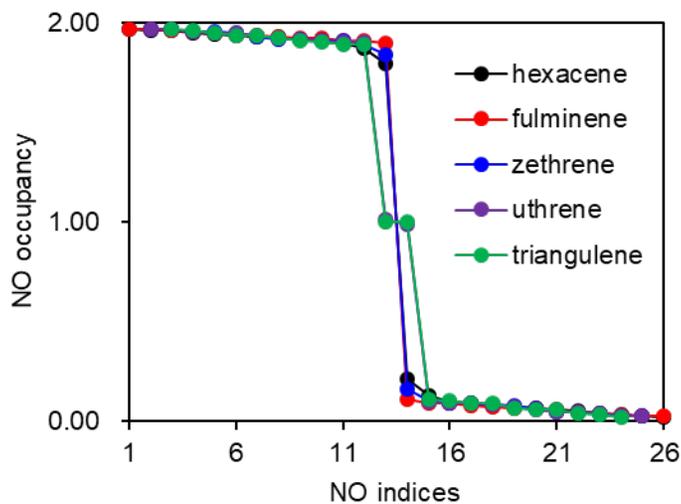

**Figure 5.** Natural orbital occupancies in 6-ring-fused PAHs. Note the offsets of the NO indices in zethrene (2), uthrene (2), and triangulene (3) for consistency in the indices of the HOMO and LUMO.

The resulting singlet–triplet gaps and $N_U$ values for singlet and triplet states are shown in Table 3. As expected, the ground states are triplet states in uthrene and triangulene and singlet states in the other systems. Based on the extrapolated energies, the singlet–triplet gap ($\Delta E_{ST}$) is largest in fulminene (~60 kcal/mol), and the gaps are similar in hexacene and zethrene. The $\Delta E_{ST}$ values of hexacene were calculated with various means of approximate CASSCF methods,[96-99] and they range from 17.1 (v2RDM-SCF/cc-pVDZ)[96] to 21.4 (DMRG/6-31+G**) kcal/mol.[97] The ASCI-SCF values of 19.0 to 19.3 kcal/mol are within this range. The numbers computed with methods that incorporate dynamical correlations (11.2, 15.0, 16.8 kcal/mol with ACI-DSRG-PT2/cc-pVDZ,[77] GAS-PDFT/6-31+G**,[99] DMRG-PDFT/6-31+G**,[97] respectively) are closer to the experimental value (~12 kcal/mol).[77,100] These results suggest that the errors in calculated values with approximate CASSCF methods can be attributed to missing dynamical correlations. In both uthrene and triangulene, the triplet state is ~10 kcal/mol more stable than the singlet state. The number of unpaired electrons in the triplet states is approximately 2.5 for all the tested systems. The $N_U$ values for singlet states of uthrene and trianglene are greater than 2.0, which means that these molecules have diradical singlet states. These states are naturally less stable than the diradical triplet states according to Hund's rule. The experimental results also support the diradical ground state (singlet or triplet) in triangulene.[101] The others have closed-shell singlet ground states. The natural orbital occupancies in the singlet states, computed with the relaxed densities, are shown in Figure 5 and reconfirm that the uthrene and triangulene singlet states are biradical, as confirmed with the MRCISD[95] or CASSCF calculations using limited active space.[94]

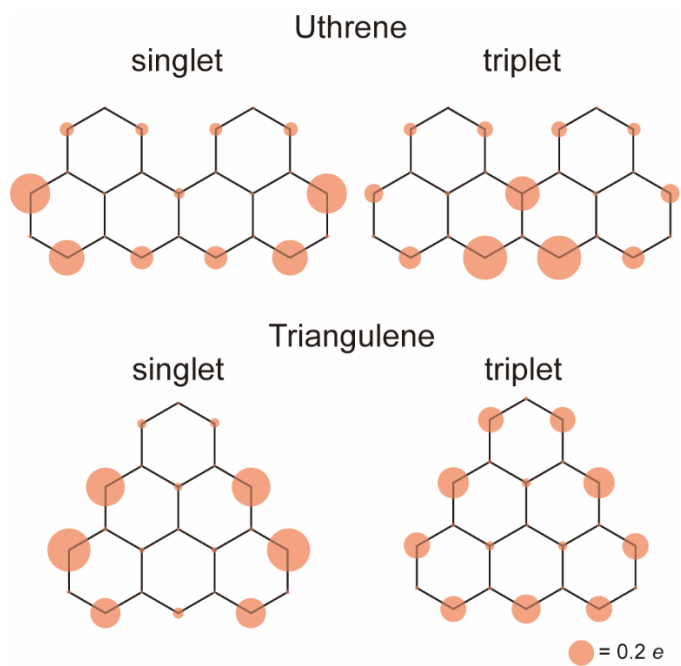

**Figure 6.** Atomic distributions of unpaired electrons in uthrene and triangulene with $N_{\text{tdet}}$ = 500000.

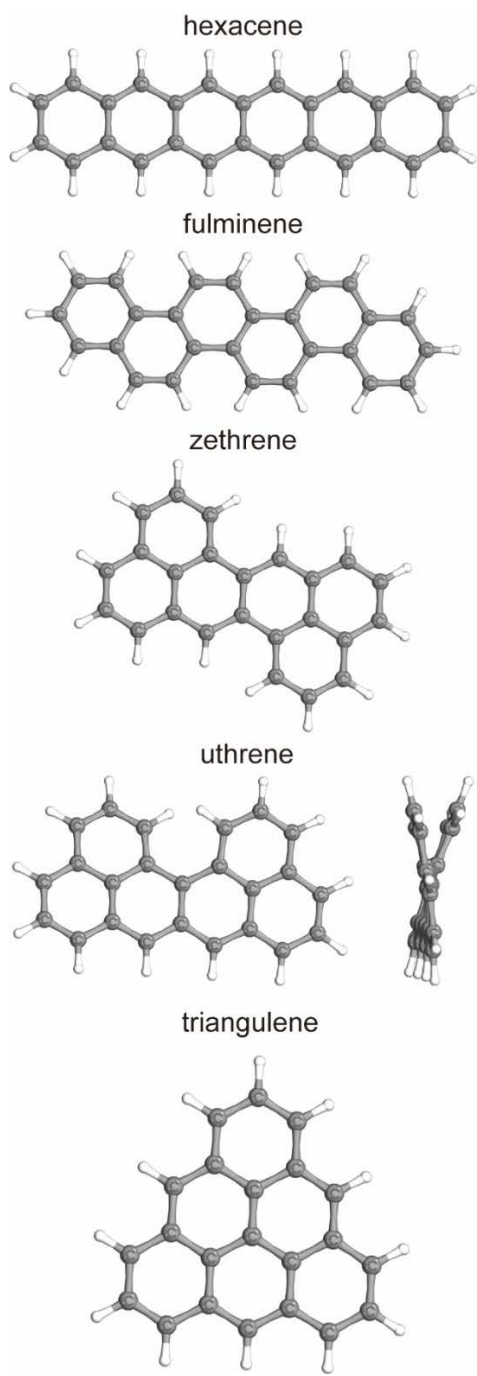

**Figure 7.** Optimized ground-state structures (singlet for hexacene, fulminene, zethrene; triplet for uthrene and triangulene) of 6-ring-fused PAHs. For utherene, the side view is also displayed. The molecular graphics are generated using the software IboView.[102,103]

We evaluated the number of unpaired electrons on each carbon atom in uthrene and triangulene by back-transformations of the unpaired electron densities into the atomic orbital basis, followed by Mulliken analysis. The results with relaxed densities at $N_{tdet} = 500000$ are displayed in Figure 6. The unpaired electrons are mainly distributed alternatively along the edge of these molecules. The atoms with large unpaired electron densities belong to the same group, as defined by Ovchinnikov's rule. In the triplet and singlet states, these electrons are placed symmetrically and nonsymmetrically, respectively.

As predicted with DFT calculations,[94] the optimized structure of uthrene exhibits geometries slightly twisted by ~20 deg, while all the other molecules are planar (Figure 7). The singlet and triplet geometries in triangulene are isosceles-like and regular-triangle-like, respectively, due to their symmetric electronic structures. In the triplet state, the optimized geometry's symmetry is slightly broken unless the CASCI (perfect) wave function is used, as we did not exploit any geometrical symmetry ($D_{3h}$). In other words, the wave function and energy accurately approximate the CASSCF functions when the optimized triplet geometry is similar to a regular triangle. The edge bond lengths and the deviations from the symmetric geometries in triangulene are shown in Figure 8. When the geometries were optimized with ASCI-SCF, ASCI-SCF-PT2, and ASCI-SCF-PT2+X, the maximum deviations from the symmetric geometry were 0.0012 Å, 0.0007 Å, and 0.0005 Å, respectively. The optimized conformations are slightly isosceles-like, and the triangular sides differ by 0.0016 Å, 0.0016 Å, and 0.0002 Å (Figure 8). Again, these results imply that minimizing the extrapolated ASCI-SCF-PT2 energy can yield reasonable approximations to the CASSCF molecular conformations.

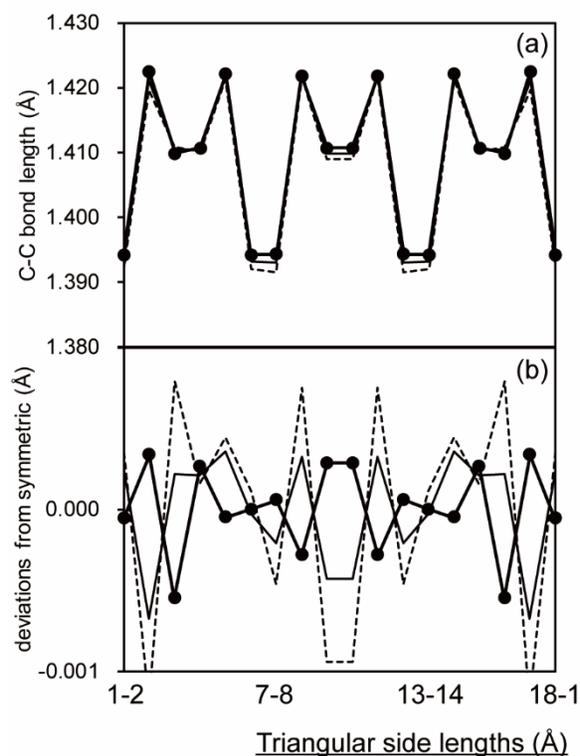

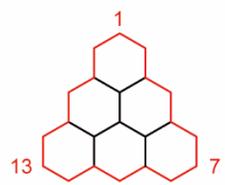

**Figure 8.** Optimized triplet geometry in triangulene. (a) C–C bond length alternation (BLA) pattern along the edge bonds and (b) their deviations from the symmetric geometry, with ASCI-SCF-PT2+X (bold solid line), ASCI-SCF-PT2 (thin solid line), and ASCI-SCF (dashed line). We defined the symmetric geometry by taking averages over the three triangular sides. The side lengths of the triangle are also shown.

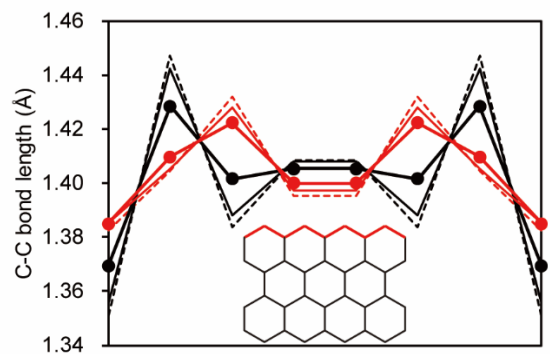

**Figure 9.** C−C BLA pattern along the zigzag π-bonds in n-periacenes under the ASCI-SCF (dashed), ASCI-SCF-PT2 (solid thin), and ASCI-SCF-PT2+X (solid bold) singlet (black) and triplet (red) optimized geometries for 4-periacene. The size of the target determinant space was $10^6$ for ASCI-SCF and ASCI-SCF-PT2 optimizations.

**Table 4.** The ASCI-SCF, ASCI-SCF-PT2, and extrapolated ASCI-SCF-PT2 singlet, triplet, and singlet–triplet gap energies at the ASCI-SCF, ASCI-SCF-PT2, and ASCI-SCF-PT2+X geometries of 4-periacene.

|  | ASCI-SCF geometry | | | ASCI-SCF-PT2 geometry | | | ASCI-SCF-PT2+X geometry | | |
|---|---|---|---|---|---|---|---|---|---|
|  | $N_{tdet}$ = 1000000 | | extrap. | $N_{tdet}$ = 1000000 | | extrap. | $N_{tdet}$ = 1000000 | | extrap. |
|  | ASCI-SCF | ASCI-SCF-PT2 |  | ASCI-SCF | ASCI-SCF-PT2 |  | ASCI-SCF | ASCI-SCF-PT2 |  |
| singlet ($E_h$) | -1373.16579 | -1373.19884 | -1373.24458 | -1373.16550 | -1373.19943 | -1373.24681 | -1373.16000 | -1373.19710 | -1373.25225 |
| triplet ($E_h$) | -1373.15784 | -1373.19306 | -1373.23363 | -1373.15779 | -1373.19337 | -1373.23588 | -1373.15547 | -1373.19403 | -1373.23815 |
| $\Delta E_{ST}$ (kcal/mol) | 4.99 | 3.63 | 6.87 | 4.84 | 3.81 | 6.86 | 2.84 | 1.92 | 8.85 |

**Case of 4-Periacene.** In our previous study using the ASCI-SCF analytical gradient,[68] we found that for 4-periacene, the ASCI-SCF-optimized singlet geometry is not the true minimum for the singlet state. The extrapolated energy of the singlet-optimized geometry was higher than that of the triplet-optimized geometry. This suggested that the CASSCF-level minimum is not the ASCI-SCF singlet geometry but rather near the ASCI-SCF triplet geometry. To find the near-CASSCF-level geometries of 4-periacene, we optimized the singlet and triplet molecular geometries with ASCI-SCF, ASCI-SCF-PT2 (both with $N_{\text{tdet}} = 10^6$), and ASCI-SCF-PT2+X. The (36$e$,36$o$) active space with full $\pi$-valence active space was employed. Four data points ($N_{\text{tdet}} = 4 \times 10^5, 6 \times 10^5, 8 \times 10^5$, and $10^6$) were utilized for extrapolation.

The total and singlet–triplet gap energies of the ASCI-SCF, ASCI-SCF-PT2, and ASCI-SCF-PT2+X geometries are shown in Table 4. We note that the total energies presented here are slightly different from those in Ref. 68, as we have extrapolated the ASCI-SCF energies instead of the ASCI energies (i.e., the extrapolation points included the orbital relaxation effects at the points with smaller $N_{\text{tdet}}$). The ASCI-SCF-PT2+X optimization relaxed the energy by 7.7 m$E_{\text{h}}$ (~5 kcal/mol) in the singlet state, and of course, the optimized energy (−1373.25225 $E_{\text{h}}$) was lower than the singlet extrapolated energy in the triplet ASCI-SCF-PT2 geometry (−1373.24464 $E_{\text{h}}$). This makes the resulting extrapolated ASCI-SCF-PT2 $\Delta E_{\text{ST}}$ 8.9 kcal/mol. This value is smaller than the v2RDM-SCF/cc-pVDZ (13.4 kcal/mol) value[96,104] but larger than the DMRG/STO-3G (5.3 kcal/mol) value.[105] This value is also larger than the recent experimental value obtained via a SQUID experiment (2.5 kcal/mol).[106] Most likely, including dynamic correlation will improve the computed $\Delta E_{\text{ST}}$ value like in the case of hexacene (see above). We note that the optimized active orbitals in the ASCI-SCF-PT2+X geometry are almost the same as those in the ASCI-SCF-PT2 geometry.

The C–C bond lengths along the longer edge in 4-periacene in various optimized geometries are shown in Figure 9. The maximum deviations between the C–C bond length in the ASCI-SCF-PT2+X and ASCI-SCF geometries were 0.019 and 0.008 Å for the singlet and triplet states, respectively. The singlet and triplet ASCI-SCF-PT2+X geometries move closer to the triplet and singlet ASCI-SCF geometries, respectively, but do not coincide, as hypothesized in our previous work[68] or as in the v2RDM-SCF/cc-pVDZ optimization.[67,96] Overall, we resolved some open questions regarding the ASCI-SCF method's performances for describing 4-periacene in our earlier work using the ASCI-SCF-PT2+X gradient. We finally note that the possible contributions of the dynamical correlation (or the electron correlation out of the active space) will probably be more significant than the ASCI-SCF-PT2+X corrections. The quantitative investigations of such dynamical correlation effects will be an intriguing target of future studies.

**Table 5.** Wall clock time for single iteration (gradient evaluation) in ASCI-SCF-PT2+X geometry optimizations. These times were measured using 18 physical cores in Intel Xeon Gold 6240 CPU (2.60 GHz). The detailed descriptions of times are in the footnote.

| | Parameters | | | | Wall clock time (secs) | | | | | | |
|---|---|---|---|---|---|---|---|---|---|---|---|
| | Active space | $N_{bas}$[a] | Largest $N_{tdet}$[b] | Number of $T^c$ | ASCI-SCF | | | ASCI-SCF-PT2 | | | Total ASCI-SCF-PT2+X[j] |
| | | | | | $t_{RDM}$[d] | $t_{OO}$[e] | $t_{list}$[f] | $t_{sort}$[g] | $t_{ASCI-PT2}$[h] | $t_{Zvec}$[i] | |
| Anthracene | (14$e$,14$o$) | 246 | 100000 | 8 162 971 | 11 | 3 | 14.7 | 18 | 81 | 371 | 2170 |
| Hexacene | (26$e$,26$o$) | 444 | 500000 | 1 433 657 291 | 60 | 20 | 442 | 355 | 2000 | 1850 | 17000 |
| 4-periacene | (36$e$,36$o$) | 584 | 1000000 | 18 274 677 825 | 140 | 75 | 3276 | 2950 | 24000 | 3380 | 71400 |

[a] Number of basis functions. [b] Largest $N_{tdet}$ employed in ASCI-SCF-PT2+X calculations. [c] Number of PT2 configurations included in the calculations with the largest $N_{tdet}$ for timing benchmark. We imposed the cutoff of $10^{-7}$ to generate these contributions. [d] time for single diagonalization and single RDM calculation. [e] time for orbital optimization in a macroiteration. [f] time for generating the list of PT2 configurations. [g] time for sorting the list of PT2 configuration. [h] time for computing ASCI-PT2 energy, RDMs, and CI derivatives. [i] time for solving the Z-vector equation. [j] time for a single ASCI-SCF-PT2+X geometry optimization step. One step includes four (hexacene, 4-periacene) or five (anthracene) ASCI-SCF-PT2 gradient computations.

**Computational Cost.** Finally, we will discuss the computational cost of the current algorithm. Table 5 shows the wall clock time for ASCI-SCF-PT2+X gradient calculations for three systems tested in this work (singlet anthracene, hexacene, and 4-periacene). On eighteen physical cores in Intel Xeon Gold 6240 CPU (2.60 GHz), an ASCI-SCF-PT2+X geometry optimization step for anthracene, hexacene, and 4-periacene took 0.6, 4.7, and 20 hours, respectively. Roughly, the wall time needed for a single optimization step increases by a factor of ~10 when the numbers of electrons and active orbitals are increased by ~10. Because we are "sampling" the electronic configurations in the CASCI space, rather than computing with a rigorous ansatz like in DMRG, it is somewhat tricky to write scaling with simple molecular parameters like $N_{act}$. Instead, we can write the computational cost using the empirical parameters, such as the number of PT2 determinants ($T$). The number of operations for evaluating ASCI-SCF-PT2 energy is $\sum_T \alpha_T$, where $\alpha_T$ is the number of the configurations in variational wave function that interact with $T$. If we assume that $\alpha_T$ is similar for all $T$, the computational cost will be proportional to the number of $T$, roughly the case for the data in Table 5.

The CASSCF calculations are not practical except for anthracene. For anthracene, a single CASCI iteration took 10 seconds. Approximating the wall time on the same CPU using the scaling of the most expensive step ($N_{det}N_{act}^4$) in the Knowles–Handy algorithm[107] yields the estimated single CASCI iteration time of ~30 and ~$10^8$ years for hexacene and 4-periacene, respectively.

Let us compare the computational cost for our ASCI-SCF-PT2+X algorithm with the other methods in the literature. The DMRG-SCF methods are considered to be the most accurate and are used for molecular geometry optimizations.[108-110] A single DMRG-SCF geometry optimization step of (20$e$,22$o$) spiropyran with the 6-31G(d) basis set took ~10 hours with $M$ = 512 using 16 cores in Intel Xeon E5-2690 CPU (2.90 GHz),[109] while $M$ denotes the bond dimension. A single

DMRG-SCF iteration in (10$e$,34$o$) indole using the aug-cc-pVTZ basis set with $M = 1000$ and (12$e$,28$o$) Cr$_2$ using the cc-pVDZ basis set with $M = 500$ required 11 hours[111] and 2 hours,[112] respectively, using 16 cores in Intel Xeon E5-2670 (2.60 GHz) and Intel Xeon E5-2667 (2.90 GHz) CPUs. The v2RDM-SCF geometry optimizations are reported to be quite efficient, as a single geometry optimization step in (26$e$,26$o$) hexacene and (38$e$,38$o$) nonacene with the cc-pVDZ basis set took ~10 mins and ~50 mins,[96] respectively, using six cores in i7-6850K CPU (3.60 GHz). The reported selected CI methods (although without geometry optimizations) require lower computational cost than our algorithm. A CIPSI calculation of (24$e$,76$o$) Cr$_2$ using the cc-pVDZ basis set with $2\times10^7$ determinants took ~14 mins using 800 physical cores at 2.70 GHz.[113] A semi-stochastic heat-bath CI calculation and ASCI-PT2 calculation of (14$e$,26$o$) F$_2$ with the cc-pVDZ basis required 5 and 49 seconds on 20 and single physical core in Intel Xeon E5-2680v2 (2.80 GHz) and Intel Xeon E5-2620v5 CPU (2.10 GHz), respectively.[42,66] Of course, we should note that these comparisons are indirect, as the calculations were performed for different systems on diverse computer architecture with various software.

The benchmark results imply that there is plenty of room for improvement in our program. The generation of contributions ($T$) to each triplet constraint is not implemented optimally (using the methods like in Ref. 42). The most expensive term is the two-particle RDM from the second-order term (Eq. 25), which requires $O(N_T N_{\text{ele}}^2)$ operations, where $N_{\text{ele}}$ is the number of electrons in the active space. One can efficiently compute the Hamiltonian diagonal element for each $T$ ($H_{TT}$ in Eq. 8) with the energy of determinants in ASCI wave function. One could similarly apply such a strategy for evaluating RDMs. Finally, the routine for computing the ASCI-related $\sigma$-term for the $Z$-vector equation should be improved as well, particularly in terms of its parallelization.

## 4. SUMMARY AND FUTURE PROSPECTS

This work has developed the analytical gradient theory for ASCI-SCF corrected with second-order perturbation theory (ASCI-SCF-PT2). To this end, we have developed the Lagrangian formalism (Z-vector equation) for the ASCI-SCF reference function. Extrapolation of the ASCI-SCF-PT2 analytical gradient with respect to the PT2 correction (ASCI-SCF-PT2+X) yields an optimized geometry that closely resembles the CASSCF conformation. This development enables geometry optimizations and molecular dynamics simulations with large active spaces (that were not tractable with the CASSCF method) at almost the CASSCF level.

The effective (relaxed) density at the ASCI-SCF-PT2 level can be obtained with the Z-vector equation solutions. The quantities related to the electronic structure, such as the number of unpaired electrons, can be obtained from the relaxed density and linearly extrapolated to obtain the near-CASSCF values, as is the case for the ASCI-SCF-PT2 energy. The source codes for the ASCI-SCF-PT2 analytical gradient are distributed in the form of patches on open-source BAGEL version 2021.02.05 at http://sites.google.com/view/cbnuqbc/codes under the GPU-v3 license.

There remains plenty of room for improvement. Of course, the current computational algorithm can be made more efficient. Multireference methods are actively used in studying excited states, so extending the theory for treating excited states is promising. This will enable the simulations that were impractical with conventional CASSCF or CASPT2, such as those for highly degenerate transition metal complexes or conical intersection dynamics. For example, one can extend the CASSCF study on the pyracylene[114] to the larger PAHs like those investigated in the previous[68] and this work. The dynamical correlation plays an important role in geometry optimizations and dynamics,[115] and the combinations of ASCI-SCF(-PT2) with the MRPT, MRCI, or MRCC methods indeed will constitute significant contributions. The *Z*-vector equation with a limited CI space can be applied for other approximate FCI methods, such as other SCI methods, RASSCF, or GASSCF, to evaluate nuclear analytical gradients of SA-RASSCF or dynamical correlation methods such as RASPT2. We will report the progress in this direction in due course.

## ACKNOWLEDGMENTS

This work was supported by the National Research Foundation (NRF) grant funded by the Korean government (MSIT) (Grant 2019R1C1C1003657) and by the POSCO Science Fellowship of POSCO TJ Park Foundation.